\documentclass[aps,prl,floatfix,showpacs,showkeys,twocolumn,superscriptaddress,nofootinbib]{revtex4-1}
\usepackage{bm}
\input{epsf}

\begin{document}

\title{Scattering of Electrons between Edge and Two-Dimensional States of a Two-Dimensional Topological Insulator and the Conductivity of the Topological Insulator Strip in a Metallic State}

\author{M.~M.~Mahmoodian}
\email{mahmood@isp.nsc.ru}
\affiliation{Rzhanov Institute of Semiconductor Physics, Siberian Branch, Russian Academy of Sciences, Novosibirsk, 630090 Russia}
\affiliation{Novosibirsk State University, Novosibirsk, 630090 Russia}

\author{M.~V.~Entin}
\email{entin@isp.nsc.ru}
\affiliation{Rzhanov Institute of Semiconductor Physics, Siberian Branch, Russian Academy of Sciences, Novosibirsk, 630090 Russia}
\affiliation{Novosibirsk State University, Novosibirsk, 630090 Russia}


\begin{abstract}
The lifetime of electrons on edge states of a two-dimensional topological insulator against the background of an allowed two-dimensional band has been determined. It has been shown that this time in the case of scattering on Coulomb impurities can be significantly larger than the mean free time of two-dimensional electrons. As a result, the conductivity of the metallic two-dimensional topological insulator strip can be determined primarily by edge states.
\end{abstract}

\maketitle

\subsection*{Introduction}

Edge states of a topological insulator are bright manifestations of its topological properties. These states cover the entire band gap of the topological insulator. The backscattering of charge carriers is forbidden in them because of topological protection. As a result, the conductance of the topological insulator in an insulating state is ballistic and nonlocal. We showed in \cite{ent-mah-mag} that these states in a number of models have a linear dispersion and extend beyond the band gap. In this work, we study the impurity scattering of carriers between edge states against the background of an allowed band and two-dimensional states. We show that, although such transitions are not topologically forbidden, the probability of Coulomb scattering is small and, correspondingly, the mean free path of carriers on edge states is anomalously large. As a result, edge states make a significant contribution to the conductivity, and the conductance of the long strip with the Fermi level in the allowed two-dimensional band can be determined by edge carriers rather than twodimensional ones.

\subsection*{Edge and two-dimensional states of a semi-infinite topological insulator}

To calculate edge and two-dimensional states of the topological insulator, we use the Volkov–Pankratov Hamiltonian \cite{vp} adapted for a two-dimensional system \cite{ent-mah-mag}. This Hamiltonian is directly applicable to a two-dimensional insulator based on a HgTe layer with a variable thickness. The thick part has a negative energy gap and corresponds to a topological insulator, whereas the thin part has a positive energy gap and corresponds to a normal insulator:
\begin{equation}\label{vp}
   H=\left(
       \begin{array}{cc}
         \Delta(y)\sigma_0 &v\bm{\sigma}{\bf k} \\
         v\bm{\sigma}{\bf k} & -\Delta(y)\sigma_0 \\
       \end{array}
     \right),
\end{equation}
where ${\bf k}=(k_x,k_y)$ is the two-dimensional momentum operator, $\sigma_0$ is the $2\times2$ identity matrix, and $\bm{\sigma}$ are the Pauli matrices. Below, except for the final expression, we set $\hbar=1$.

Further, we consider a stepwise dependence of the gap $\Delta(y)=\Delta_1 \theta(-y)+\Delta_2 \theta(y)$, where $\Delta_1>0$ and $\Delta_2<0$; i.e., the $y<0$ half-plane is the normal insulator and the $y>0$ half-plane is the topological insulator.

Hamiltonian (\ref{vp}) with the constant gap $\Delta_{1,2}$ has the eigenfunctions
\begin{eqnarray}\label{wfTI}
&&\Psi^{(1,2)}_{{\bf k},\sigma}=\zeta^{(1,2)}_{{\bf k},\sigma}e^{ik_x x+ik_y y},\\
&&\zeta^{(1,2)}_{{\bf k},+1}=(1,0,0,\alpha^{(1,2)}_{{\bf k},+1}),\\
&&\zeta^{(1,2)}_{{\bf k},-1}=(0,1,\alpha^{(1,2)}_{{\bf k},-1},0),\\
&&\alpha^{(1,2)}_{{\bf k},\sigma}=\frac{v(k_x+i\sigma k_y^{(1,2)})}{E+\Delta_{1,2}}\label{alpha}
\end{eqnarray}
with the energies
\begin{eqnarray}\label{EcvI}
E=\mu\sqrt{\Delta_{1,2}^2+v^2k^2}.
\end{eqnarray}
Here, $\sigma=\pm1$ is the spin index and the index $\mu=\pm1$ corresponds to positive and negative energies, respectively; energies for real ${\bf k}$ are in the conduction and valence bands, respectively. The eigenfunctions of Hamiltonian (\ref{vp}) localized near $y=0$ consist of damping waves with $k_y^{(1,2)}=-i\lambda_{1,2}$, $\lambda_1>0$ and $\lambda_2<0$ for $y<0$ and $y>0$, respectively. Using Eq. (\ref{EcvI}), we find $\lambda_{1,2}^2=(\Delta_{1,2}^2-E^2)/v^2+k_x^2$.

\begin{figure}[ht]
\leavevmode\centering{\epsfxsize=6.5cm\epsfbox{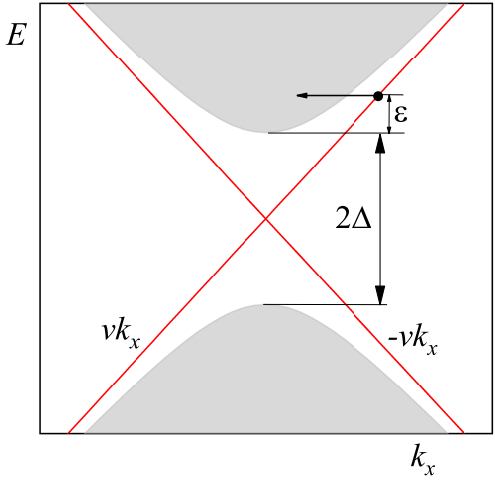}}
\caption{Energy diagram of the two-dimensional topological insulator with $\Delta_1=-\Delta_2=\Delta$. Edge states overlap in energy with two-dimensional ones. Impurity scattering occurs between edge and two-dimensional states. At transitions near the bottom of two-dimensional band, the momentum in the edge state is much lower than the momentum in the two-dimensional state.}\label{fig1}
\end{figure}

From the condition of continuity $\Psi^{(1)}_{{\bf k},\sigma}|_{y=0}=\Psi^{(2)}_{{\bf k},\sigma}|_{y=0}$, we determine the energies of edge states $E=\epsilon_{k_x,\sigma}\equiv\sigma vk_x$, parameters $\lambda_{1,2}=\Delta_{1,2}/v$, and localized edge eigenfunctions
\begin{eqnarray}\label{wfEd}
\psi_{k_x,\sigma}=\frac{C\chi_{k_x,\sigma}}{\sqrt{L_x}}&&e^{ik_x x}\times
\Bigg\{
\begin{array}{ccc}
  e^{\frac{\Delta_1}{v}y},~ & \mbox{ïðè} & y<0, \\
  e^{\frac{\Delta_2}{v}y}, & \mbox{ïðè} & y>0,
\end{array}\\
&&\chi_{k_x,+1}=(1,0,0,1),\nonumber\\
&&\chi_{k_x,-1}=(0,1,-1,0),\nonumber\\
&&C=\frac12\sqrt{\frac{\Delta_1|\Delta_2|}{v\left(\Delta_1+|\Delta_2|\right)}}.\nonumber
\end{eqnarray}
Here and below, $L_x$ è $L_y$ are the dimensions of the sample in the $x$ and $y$ directions, respectively. The velocities of electrons on localized states are $\sigma v$. These localized solutions are the only solutions in the energy range $|E|<|\Delta_2|$. However, they exist for all energies (see Fig.~\ref{fig1}).

Solutions delocalized in the $y$ direction exist at $E>\Delta_1$ ($\Delta_1>|\Delta_2|$). They can be specified by the continuous wave vector $k_y>0$:
\begin{eqnarray}\label{wf2d0}
&&\Psi_{{\bf k},\sigma}=\frac{e^{ik_x x}}{\sqrt{L_x}}\times\\
&&\Bigg\{
\begin{array}{ccc}
  C^{(1+)}_{{\bf k},\sigma}\zeta^{(1)}_{{\bf k},\sigma}e^{ik_y y}+C^{(1-)}_{{\bf k},\sigma}\zeta^{(1)}_{{\bf k},-\sigma}e^{-ik_y y}, & \mbox{ïðè} & y<0, \\
  C^{(2+)}_{{\bf k},\sigma}\zeta^{(2)}_{{\bf k},\sigma}e^{ik_y y}+C^{(2-)}_{{\bf k},\sigma}\zeta^{(2)}_{{\bf k},-\sigma}e^{-ik_y y}, & \mbox{ïðè} & y>0.
\end{array}\nonumber
\end{eqnarray}

Let the incident and reflected waves exist on the side of the normal insulator ($y<0$) and the transmitted wave exist on the side of the topological insulator ($y>0$):
\begin{eqnarray}\label{wf2d}
&&\Psi_{{\bf k},\sigma}=\frac{e^{ik_x x}}{\sqrt{L_x}}\times\\
&&\Bigg\{
\begin{array}{ccc}
  C^{(1+)}_{{\bf k},\sigma}\zeta^{(1)}_{{\bf k},\sigma}e^{ik_y y}+C^{(1-)}_{{\bf k},\sigma}\zeta^{(1)}_{{\bf k},-\sigma}e^{-ik_y y}, & \mbox{ïðè} & y<0, \\
  C^{(2+)}_{{\bf k},\sigma}\zeta^{(2)}_{{\bf k},\sigma}e^{ik_y y},~~~~~~~~~~~~~~~~~~~~~~~~~~~ & \mbox{ïðè} & y>0.
\end{array}\nonumber
\end{eqnarray}

It follows from the continuity of the wavefunction at the interface that
\begin{eqnarray}\label{wf2dcont}
C^{(1\pm)}_{{\bf k},\sigma}=b^{(\pm)}_{{\bf k},\sigma}C^{(2)}_{{\bf k},\sigma},\\
b^{(\pm)}_{{\bf k},\sigma}=\mp\frac{\alpha^{(2)}_{{\bf k},\sigma}-\alpha^{(1)*}_{{\bf k},\sigma}}
{\alpha^{(1)}_{{\bf k},\sigma}-\alpha^{(1)*}_{{\bf k},\sigma}},
\end{eqnarray}
where $\alpha^{(1,2)}_{{\bf k},\sigma}$ is given by Eq. ({\ref{alpha}}) with $E=\mu\varepsilon_k\equiv\mu\sqrt{\Delta_2^2+v^2k^2}$.

The normalization of the wavefunction (\ref{wf2d}) gives
\begin{eqnarray}\label{wf2dnorm}
C^{(2)}_{{\bf k},\sigma}=\Bigg[\frac{L_y}{2}\bigg\{\left(\left|b^{(+)}_{{\bf k},\sigma}\right|^2+\left|b^{(-)}_{{\bf k},\sigma}\right|^2\right)\left(1+\left|\alpha^{(1)}_{{\bf k},\sigma}\right|^2\right)+\nonumber\\
+1+\left|\alpha^{(2)}_{{\bf k},\sigma}\right|^2\bigg\}\Bigg]^{-1/2}.
\end{eqnarray}

\subsection*{Lifetime on edge states of the two-dimensional topological insulator}

Electrons with a given spin on an edge state move in one direction. The only possible mechanism of scattering between edge states is backscattering. However, the conservation of spin forbids it in the absence of breaking of the time reversal symmetry. However, edge states are not isolated from two-dimensional states in the presence of elastic scattering. Being in the potential of impurities, electrons can pass from edge states to two-dimensional states and back. To determine the lifetime of the electron $\tau$ on an edge state, we represent the probability of scattering from the edge state to the two-dimensional state in the Born approximation in the form
\begin{equation}\label{w}
W_{k_x,\sigma;\mu,{\bf k}',\sigma}=\frac{2\pi}{\hbar^2}\sum\limits_i\int\left|V_{k_x,\sigma;\mu,{\bf k}',\sigma}\right|^2\delta(vk_x-\mu\varepsilon_{{\bf k}'}).
\end{equation}
Here, $V({\bf q})=2\pi e^2/\kappa q$ is the two-dimensional Fourier transform of the potential of an unscreened charged impurity, $e$ is the elementary charge, and $\kappa$ is the dielectric constant of the medium (screening is neglected). Summation is performed over the impurity number $i$.

The matrix element of the impurity potential has the form
\begin{equation}\label{me}
V_{k_x,\sigma;\mu,{\bf k}',\sigma}=\int\psi^{*}_{k_x,\sigma}V({\bf r-r}_i)\Psi_{{\bf k}',\sigma}d{\bf r},
\end{equation}
where $V({\bf r})=\int d{\bf q}V({\bf q})e^{i{\bf qr}}/(2\pi)^2$. In view of Eqs. (\ref{wfEd}) and (\ref{wf2d})
\begin{eqnarray}\label{me1}\nonumber
&&V_{k_x,\sigma;\mu,{\bf k}',\sigma}=CC^{(2)}_{{\bf k},\sigma}\frac{e^2}{\kappa qL_x}\delta\left(q_x+k_x'-k_x\right)\times\\
&&e^{-iq_y y_i}\Bigg[\frac{b^{(+)}_{{\bf k},\sigma}\left(1+\alpha^{(1)}_{{\bf k},\sigma}\right)}{\frac{\Delta_1}{v}+i\left(q_y+k_y'\right)}+
\frac{b^{(-)}_{{\bf k},\sigma}\left(1+\alpha^{(1)*}_{{\bf k},\sigma}\right)}{\frac{\Delta_1}{v}+i\left(q_y-k_y'\right)}-\\
&&-\frac{1+\alpha^{(2)}_{{\bf k},\sigma}}{\frac{\Delta_2}{v}+i\left(q_y+k_y'\right)}\Bigg].\nonumber
\end{eqnarray}

We calculate the matrix element under the assumption that $\Delta_1=-\Delta_2=\Delta$ and that the energy of the electron is near the bottom of the two-dimensional band, $E-\Delta\equiv\epsilon\ll \Delta$. The corresponding momentum transfer is $q\gg k$. These assumptions strongly simplify the answer. Then, in view of Eq. (\ref{me1}), the probability of scattering has the form
\begin{eqnarray}\label{w1}
W_{k_x,\sigma;\mu,{\bf k}',\sigma}=\frac{3\pi n_i}{\kappa^2}\left(\frac{ev}{2\Delta}\right)^4k_y'^2\delta(vk_x-\mu\varepsilon_{{\bf k}'}),
\end{eqnarray}
where $n_i$ is the concentration of impurities. Integrating over the momentum ${\bf k}'$, we obtain the inverse lifetime of the electron on the edge state
\begin{eqnarray}\label{tau}
\frac{1}{\tau_{ed}}=\frac{3\pi^2n_ie^4\epsilon}{8\kappa^2\Delta^2}.
\end{eqnarray}

The inverse transport time of scattering of twodimensional electrons on charged impurities is $1/\tau_2=\pi^2 e^4n_i/\kappa^2\epsilon_F$, where $\epsilon_F$ is the Fermi energy measured from $\Delta$. As compared to $1/\tau_2$, the found probability of scattering of edge electrons in two-dimensional states includes a small parameter $\tau_2/\tau_{ed}=(3/8)(\epsilon_F/\Delta)^2$. The suppression of scattering from edge states to two-dimensional ones as compared to Coulomb scattering of two-dimensional electrons is partially due to a high transverse momentum of edge states (and the corresponding transverse part of the energy), which reduces the Coulomb interaction. An additional suppression factor $\sim \epsilon_F/\Delta$ is due to the smallness of the wavefunction of two-dimensional electrons near the interface $y=0$. This smallness exists because the gap jump for low-energy twodimensional electrons (at $E\to\Delta$) with nonzero $k_x$ serves as an infinite barrier (if $k_x=0$, the barrier is nonreflecting but only at the single point $k_x=0$).

At $\epsilon_F=1$ meV and $\Delta=7.5$ meV corresponding to the thicknesses of layers 5.6 è 7 nm \cite{qi}, this parameter is 0.0067. As a result, edge electrons can have significant mean free paths and, correspondingly, make a potentially large contribution to the conductivity of the medium, although they occupy a small area.

\subsection*{Conductance of a topological insulator strip}

Since the one-dimensional conductivity through the edge state is high, the conductance of the sample even in a metallic state can be determined by its edge rather than interior. We consider a topological insulator strip with the dimensions $L_x\gg L_y\gg v/\Delta$ with a degenerate electron gas. The conductance of the inner region in such strip can be estimated as $\Sigma_2=(e^2/2\pi\hbar) k_Fl_2L_y/L_x$. At the same time, the conductance of the edge state is $\Sigma_1=4(e^2/2\pi \hbar) l_1/L_x$. Consequently, $\Sigma_1/\Sigma_2 =4l_1/(l_2k_FL_y)$, and $\Sigma_1>\Sigma_2$ at $L_yk_F<4l_1/l_2$.

\subsection*{Discussion}

The results have been obtained within the Volkov–Pankratov model \cite{vp}. This circumstance is insignificant because the reason for the smallness of the probability of scattering is a high characteristic momentum of edge states as compared to two-dimensional electrons, which is insensitive to a model. It is noteworthy that developed fluctuations of the thickness in a system with a thickness close to a critical value of 6.3 nm cover the entire sample, creating the developed network of internal edge states. This can make the edge contribution decisive.

This work was supported by the Russian Foundation for Basic Research (project no. 17-02-00837).

\end{document}